\documentclass[preprint,12pt]{elsarticle}

\usepackage{amssymb}
\usepackage{amsmath}
\usepackage{physics}
\usepackage{pgfplots}
\usepackage{pgfkeys}
\pgfplotsset{compat=1.14} 
\newenvironment{customlegend}[1][]{%
        \begingroup
        \csname pgfplots@init@cleared@structures\endcsname
        \pgfplotsset{#1}%
    }{%
        \csname pgfplots@createlegend\endcsname
        \endgroup
    }%

    \def\addlegendimage{\csname pgfplots@addlegendimage\endcsname}
\pgfplotsset{
cycle list={%
{draw=black,mark=star,solid},
{draw=green, mark=square,solid},
{draw=black,mark=+,solid},
{draw=red,mark=o,solid},}}

\journal{Physics Letters B}

\begin{document}

\begin{frontmatter}



\title{Deviation from the First Law of Thermodynamics for Particle-like Quantum Kerr-de Sitter Black Holes}


\author[label1]{Dripto Biswas}
\ead{dripto.biswas@niser.ac.in}
\author[label1]{Jyotirmaya Shivottam}
\ead{jyotirmaya.shivottam@niser.ac.in}

\address[label1]{School of Physical Sciences, \\ National Institute of Science Education and Research,\\
Jatni - 752050, Odisha, India.}

\begin{abstract}
The goal of this paper is to extend the particle-like quantization scheme presented in \citet{silk} (2020), to extremal Kerr-de Sitter black holes in four spacetime dimensions, thereby obtaining various quantized parameters, like the black hole mass and angular momentum, consistent with existing results, in the proper limits. Moreover, we show numerically, that for such extremal quantum black holes, there is a root mean square deviation from the First Law of black hole thermodynamics, of the order $\mathcal{O}(\Lambda^{0.45 \pm 0.01})$, where $\Lambda$ denotes the Cosmological Constant.
\end{abstract}



\end{frontmatter}


\section{\label{sec:intro}Introduction}
The thermodynamics of classical Kerr-de Sitter (KdS) black holes has been investigated in \citep{law1deriv,zhang,sekiwa}, where it was shown, that they follow the First Law of black hole thermodynamics (see Sec. \ref{sec:firstlaw} for more details). In a more recent paper \citep{silk}, the authors have considered a particle-like picture of Kerr black holes in asymptotically flat backgrounds, and also verified that the First Law is satisfied for large quantum numbers, as expected from the Correspondence Principle. As mentioned in \citep{silk}, such a description may be used for studying Planck-scale relics.

In this article, we show that the particle-like quantization scheme described in \citep{silk}, when extended to de-Sitter backgrounds, only admits finitely many excitation states ($n$) for a fixed $\Lambda$. Furthermore, there is a Root Mean Square (RMS) deviation from the First Law by an order, $\mathcal{O}(\Lambda^{0.45 \pm 0.01})$ (Sec. \ref{subsec:plots}), which vanishes for $\Lambda\rightarrow 0$ (consistent with existing literature) \footnote{We hasten to remind the reader, that the vanishing of RMS deviation does not imply, that the First Law is satisfied for all $n$. Indeed, it is still violated for small $n$ even in this case ($\Lambda=0$) and is only satisfied in the large $n$ limit, as expected from the Correspondence Principle.}. 

\section{\label{sec:kds}Kerr-de Sitter Spacetime}
The Kerr-de Sitter black hole (KdS) in $(3+1)$ dimensional spacetime can be described by the following line element in Boyer-Lindquist coordinates $(t,r,\theta,\phi)$ \citep{,zhang,sekiwa},
\begin{equation}
\dd{s^2} = -\frac{\Delta_r}{\rho^2}\left(\dd{t} - \frac{a^2\sin^2\theta}{\Xi}\dd\phi\right)^2 + \frac{\rho^2}{\Delta_r}\dd{r^2} + \frac{\rho^2}{\Delta_\theta}\dd\theta^2 + \frac{\Delta_\theta \sin^2\theta}{\rho^2}\left(a \dd{t} - \frac{r^2 + a^2}{\Xi}\dd\phi\right)^2,
 \label{metric}
\end{equation}
where, 
\begin{equation}
    \begin{split}
        \rho^2 = r^2 + a^2 \cos^2\theta,\quad\Delta_\theta = 1 + \frac{\Lambda}{3}a^2 \cos^2\theta \\
        \Xi = 1 + \frac{\Lambda}{3}a^2,\quad
        \Delta_r = (r^2 + a^2)\left(1-\frac{\Lambda}{3}r^2\right) - 2m r.
    \end{split}
\end{equation}
We also know, that the KdS black hole mass ($M$) is given by $M = m/\,\Xi^2$. Throughout this paper, we have used Planck units ($c = \hbar = G = k_B = 1)$.
\subsection{Horizons of the KdS spacetime}
\label{subsec:hors}
For the purposes of this article, we shall consider the inner (Cauchy) horizon (CH) and the outer (event) horizon (EH) denoted by $r_-$ and $r_+$ respectively. We assume that there are 3 positive roots of $\Delta_r(r)=0$, corresponding to the CH, EH and the cosmological horizon $r_c$. Using, $\Delta_r(r_-) = \Delta_r(r_+) = 0$, one can write, after slightly long but simple algebra, the following,
\begin{align}
        a^2 &= \frac{r_- r_+\left[1 - \frac{\Lambda}{3} (r_+^2 + r_-^2 + r_- r_+)\right]}{1 + \frac{\Lambda}{3}r_- r_+},\nonumber\\
       2m &= \frac{(r_- + r_+)(r_-^2 + a^2)(r_+^2 + a^2)}{r_- r_+(r_+^2 + r_-^2 + r_- r_+ + a^2)},\nonumber\\
       \Xi &= \frac{r_- r_+(r_+^2 + r_-^2 + r_- r_+ 2a^2 )-a^4}{r_- r_+(r_+^2 + r_-^2 + r_- r_+ + a^2)}.
       \label{horquant}
\end{align}
We shall use these expressions extensively in the rest of the paper. Moreover, for simplicity, we shall study only the extremal KdS black hole characterised by $r_+ = r_- = \Gamma$ (say) henceforth. The various quantities in (\ref{horquant}) will be investigated in the following sections.

\section{\label{sec:qkds}Quantum KdS Black Holes}
In this section, we shall follow the quantization scheme outlined in \citep{silk}, where they have considered a particle-like picture of a black hole, first discussed in \citep{particleRN} for a Reissner-Nordström black hole. Our primary assumption is that the EH should not be smaller than the associated Compton wavelength of the black hole, i.e. $r_+ = \frac{n}{M} = \Gamma$, where $n \in \mathbb{N}$ (in Planck units) denotes the excitation state of the quantum black hole. Substituting this value in the expression for $a^2$ in (\ref{horquant}), along with the extremality assumption, we get,
\begin{equation}\label{quanta}
a^2 = \frac{\Gamma^2(1 - \Lambda \Gamma^2)}{1 + \frac{\Lambda}{3}\Gamma^2} = \frac{n^2}{M^2}\left(\frac{1 - \frac{\Lambda n^2}{M^2}}{1 + \frac{\Lambda n^2}{3 M^2}}\right).
\end{equation}
Therefore, the angular momentum $J_H = Ma$ of the black hole EH is given by,
\begin{equation}\label{angmom}
J_H = n\left(\frac{1 - \frac{\Lambda n^2}{M^2}}{1 + \frac{\Lambda n^2}{3 M^2}}\right)^\frac{1}{2}.
\end{equation}
Clearly, this gives the correct result, as assumed in \citep{silk} (for the extremal case) when $\Lambda=0$. Other relevant quantities in \citep{silk,zhang,sekiwa} can be easily calculated following the quantization scheme, described above, and are all consistent with their asymptotically flat analogues. We shall skip those calculations and move on to a discussion of the First Law of black hole thermodynamics for the extremal case, in the following section.

Using the expressions for $m$ and $\Xi$ in (\ref{horquant}), we can also obtain the following expression for $M$,
\begin{equation}\label{massexpr}
     \frac{n (3 M^2 + n^2 \Lambda)}{3 (M^2 + n^2 \Lambda)^2}=1.
\end{equation}
Unfortunately, (\ref{massexpr}) cannot be analytically simplified further. However, one may again note, that for $\Lambda=0$, (\ref{massexpr}) gives, $M = \sqrt{n}$, which is the result obtained in Eq. (3) of \citep{silk}, for the extremal case.

\section{\label{sec:firstlaw}First Law of thermodynamics}
Before stating the Law, we mention that the angular velocity of the KdS horizons are given by \citep{zhang,sekiwa} as,
\begin{equation}\label{omega}
    \tilde{\Omega} = \frac{a\Xi}{\Gamma^2 + a^2}, ~ \Omega_{\infty} = \frac{\Lambda}{3}a,
\end{equation}
while the black hole EH angular velocity is given by $\Omega_H=\tilde{\Omega} - \Omega_\infty$.

The general form of the First Law of thermodynamics for uncharged black holes, may now be stated as\footnote{Ideally we should write, $\delta M = T_H\delta S_H + \Omega_H\delta J_H + V_H \delta P_H$, where $V_H$ is the horizon volume and $P_H = -\Lambda/8\pi$. Since in our case, $\Lambda$ is a constant, we simply ignore this term.}, 
\begin{equation}\label{firstlaw}
    \delta M = T_H\delta S_H + \Omega_H\delta J_H,
\end{equation}
where $S_H$ is the horizon entropy, given by $S_H = \pi (\Gamma^2 + a^2)/\,\Xi$, and $T_H$ is the Hawking temperature \citep{hawkrad} of the EH, given by,
\begin{equation}\label{Th}
    T_H = \frac{1}{4\pi}\frac{\partial_r\Delta_r |_{\Gamma}}{\Gamma^2 + a^2} = 0,
\end{equation}
as expected for an extremal black hole. (\ref{Th}) is straightforward to check by substituting the values of the various quantities in $\partial_r\Delta_r$ and evaluating at $\Gamma = r_+ = r_-$. Therefore, to check (\ref{firstlaw}), we simply need to consider the second term on the RHS. In the following subsection, we numerically calculate both sides of (\ref{firstlaw}) for varying $n$.

\subsection{\label{subsec:numan}Numerical Analysis}
Since the analytical expressions of the quantities, in the First Law, could only be obtained as complicated identities or expressions (for e.g. $M$ in (\ref{massexpr})), we proceed to numerically calculate all relevant quantities for a given $\Lambda$ and over all \textit{allowed} values of $n$\footnote{As discussed in the following subsection, the range of \textit{allowed} $n$ refers to those values of $n$ starting from $1$, for which we have $J_H \in \mathbb{R}^+$.}. Furthermore, we confirm, that (\ref{firstlaw}) is not exactly valid for this quantum KdS black hole model (although it is seen to be valid for $\Lambda \rightarrow 0,\:n \rightarrow \infty$, as verified in \citep{silk}). To show this, we plotted $\delta M= M_{n+1}-M_n$ and compared it with $\Omega_H^n \delta J^n_H$ (superscript denotes the excitation state) obtained using (\ref{massexpr}), (\ref{angmom}) and (\ref{omega}).

\begin{figure}
\includegraphics[width=0.49\textwidth]{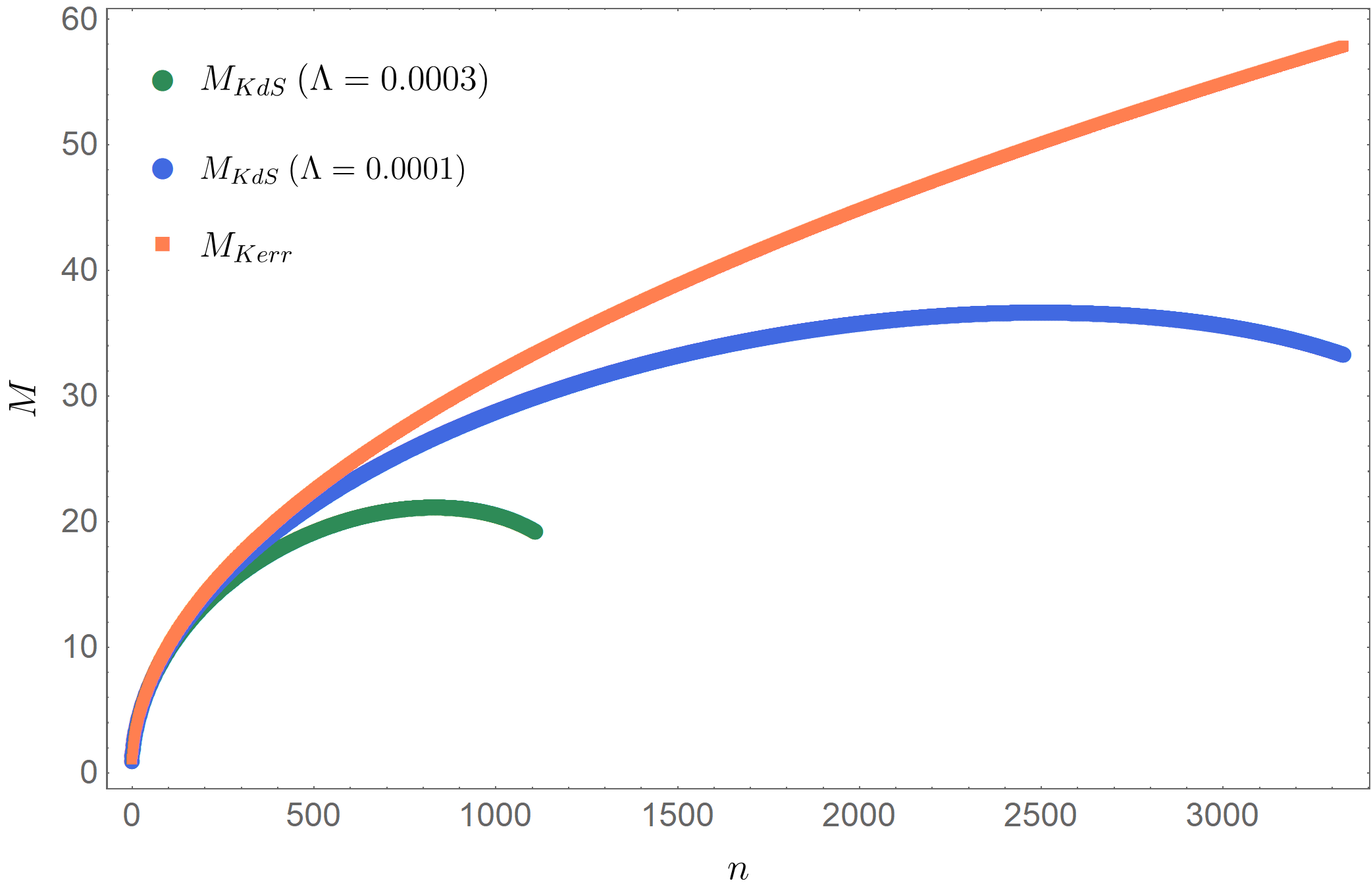}
\hspace*{\fill}
\includegraphics[width=0.49\textwidth]{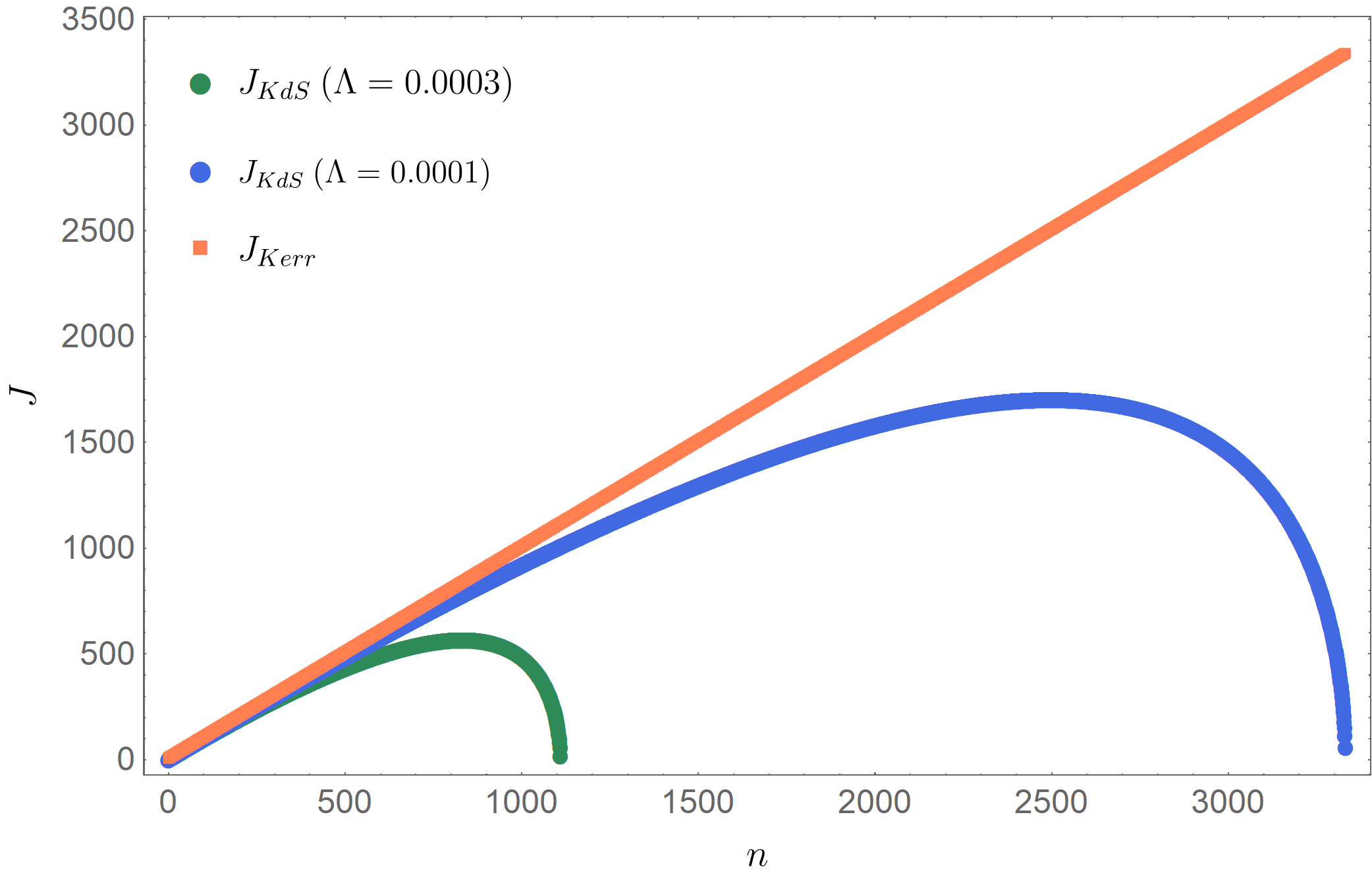}
\caption{(Left) - Plots depicting the mass, $M_{KdS}(\Lambda)$, for two different values of $\Lambda$ and the extremal Kerr black hole mass, ($M_{Kerr}$) \citep{silk}. (Right) - Similar plots for the angular momentum, $J_{KdS}(\Lambda)$, of an extremal quantum Kerr black hole and its analogue, $J_{Kerr}$ ($\Lambda=0$).}\label{fig:MJ}
\end{figure}

\subsection{\label{subsec:plots}Plots and Results}
In Fig. \ref{fig:MJ}, we plot the masses and angular momenta corresponding to the KdS black hole excitation states, $n$. The orange curve denotes the extremal Kerr analogues of the respective quantities, found in \citep{silk}. There is always a maximal angular momentum attained for finite $\Lambda$, after which $J_H$ decreases with increasing $n$. We observe that the mass (energy), $M$, follows the $\sqrt{n}$ behaviour for small $n$, but deviates significantly for larger $n$, also demonstrating a maximum possible mass (energy state), which is attained for some finite $n$, with $\Lambda \neq 0$. However, for $\Lambda \rightarrow 0$, we see, from the plots in Fig. \ref{fig:MJ}, that $M \rightarrow \sqrt{n}$ and $J_H \rightarrow n$, as expected from their analytic expressions (\ref{angmom}) and (\ref{massexpr}).

\begin{figure}
\begin{tikzpicture}
    \begin{customlegend}[legend columns=3,legend style={align=right,draw=none,column sep=2ex},legend entries={$\hspace{8em}\color{red}{\bullet}\quad\color{black}{\delta M}$, $\Omega\delta J$, $\delta M - \Omega\delta J$}]
        \addlegendimage{red, mark=o*, only marks}
        \addlegendimage{green, mark=square*, only marks}  
        \addlegendimage{blue, mark=diamond*, only marks}
    \end{customlegend}
\end{tikzpicture}

\includegraphics[width=0.49\textwidth]{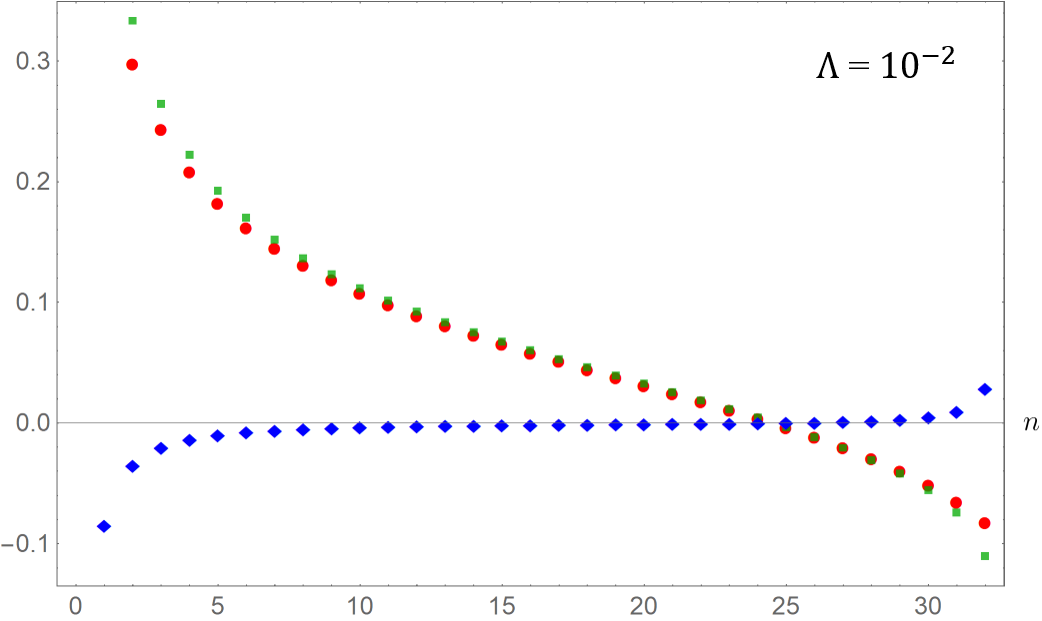}
\hspace*{\fill}
\includegraphics[width=0.49\textwidth]{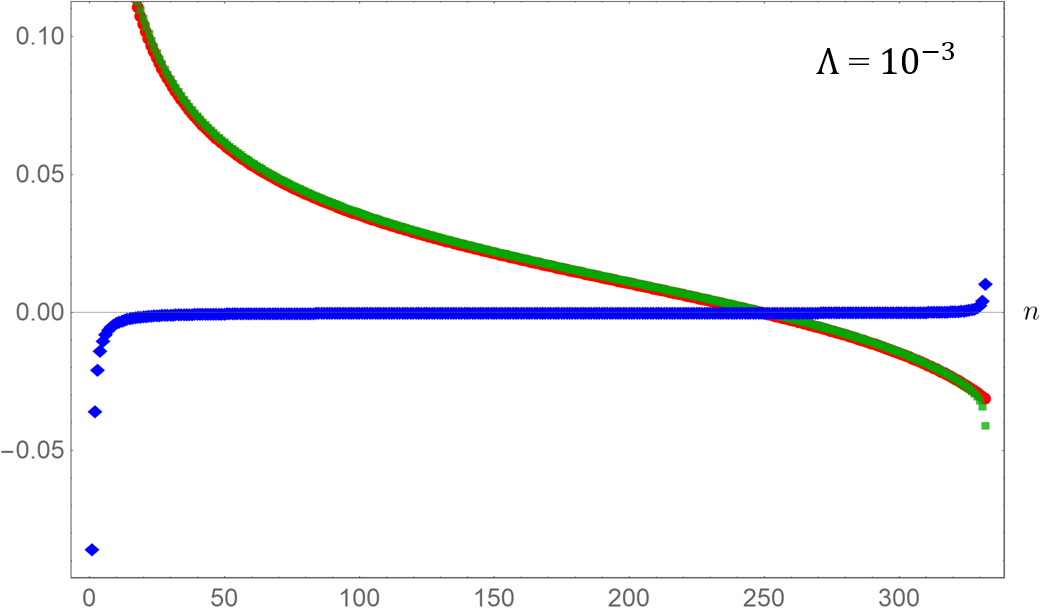}

\includegraphics[width=0.49\textwidth]{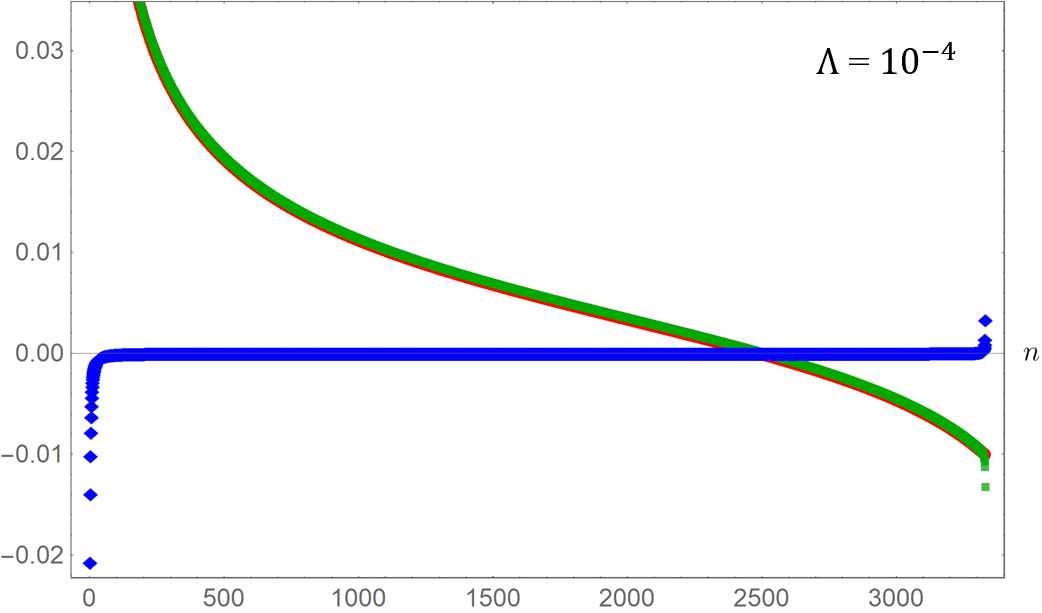}
\hspace*{\fill}
\includegraphics[width=0.49\textwidth]{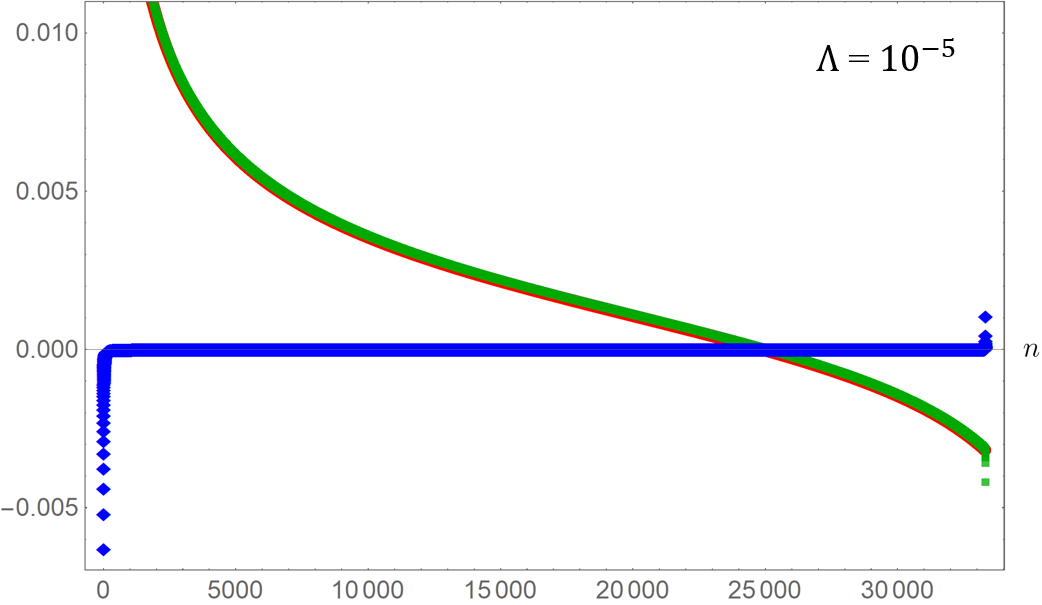}
\caption{Plots, showing the variations of $\delta M$ and $\Omega\delta J$ in $n$, for four values of $\Lambda$. The difference is depicted by the blue curve. The maximum deviation occurs near the boundaries of \textit{allowed} $n$.}\label{fig:law1plots}
\end{figure}

\begin{figure}
\centering
\includegraphics[width=0.75\textwidth]{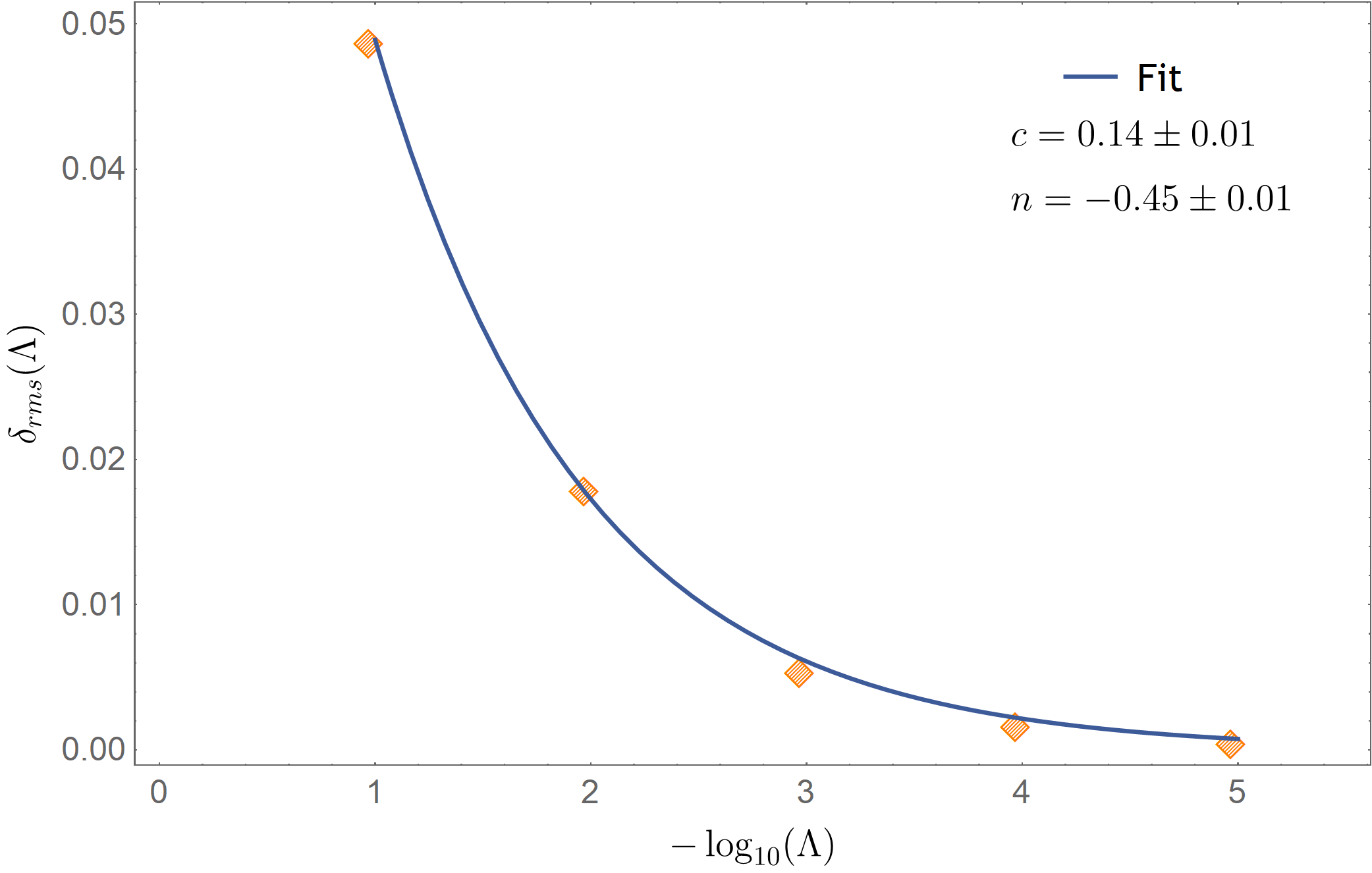}
\caption{Best-fit curve for the trial functions $f(x) = c\:10^{n x}$ with fit-parameters $c$ and $n$.}\label{fig:fit}
\end{figure}

In Fig. \ref{fig:law1plots}, we plot $\delta M$ and $\Omega\delta J$, from (\ref{firstlaw}) (red and green curves respectively), along with their difference (blue curve). It can be seen, that the largest deviations from (\ref{firstlaw}) are near the boundary values of the allowed excitation ($n$) states. Finally, we calculated the Root Mean Square (RMS) deviation from the first law $(\delta_{RMS}(\Lambda))$ and fit the $\delta_{RMS}(\Lambda)$ vs $-log_{10}(\Lambda)$ plot in Fig. \ref{fig:fit}, using a trial function of the form $f(x) = c\:10^{n x}$ (where $c$ and $n$ are curve-fit parameters).
From the values obtained in the fit, one can calculate that $\delta_{RMS}(\Lambda)$ is of the order, $\mathcal{O}(\Lambda^{0.45 \pm 0.01})$.

\section{\label{sec:results}Conclusion}
In the initial sections of this paper, we investigated the particle-like picture of an extremal Kerr black hole in de-Sitter background, essentially extending the work of \citep{silk}. Novelty of this article lies in estimating the RMS deviation from the First Law, $\delta_{RMS}(\Lambda)$ (which is of order $\mathcal{O}(\Lambda^{0.45 \pm 0.01})$), due to quantization on a de-Sitter background. Interestingly, this deviation vanishes for both quantum Kerr black holes ($\Lambda=0$) \citep{silk}, as well as the classical case, in normal KdS spacetime \citep{zhang,sekiwa} and is therefore consistent with previous results in the appropriate limits. From the plots, it is evident, that the RMS deviation arises due to the finitely many \textit{allowed} $n$ for any $\Lambda \neq 0$. To the best of our knowledge, there has been no conclusive investigation as to the physical reasons behind the existence of only finitely many energy states. A completely consistent theory of quantum gravity might be able to re-affirm (or negate) the deviation results obtained here. Nevertheless, this is an interesting observation requiring further investigations.


\bibliographystyle{elsarticle-num-names}
\bibliography{refs}

\end{document}